\documentclass[journal=ancham,manuscript=article]{achemso}

\usepackage[version=3]{mhchem} 
\usepackage[graphicx]{}
\usepackage[textcomp]{}

\author{Cristian G. Arsene}
\affiliation[Physikalisch-Technische Bundesanstalt, D-38116 Braunschweig, Germany]{Physikalisch-Technische Bundesanstalt (PTB)}
\email{christian.arsene@ptb.de}

\author{J\"urgen Kratzsch}
\affiliation[Institute for Laboratory Medicine, Clinical Chemistry and Molecular Diagnostics, University of Leipzig, D-04103 Leipzig, Germany]{Institute for Laboratory Medicine, Clinical~Chemistry and Molecular Diagnostics, University of Leipzig}

\author{Andr\'e Henrion}
\affiliation[Physikalisch-Technische Bundesanstalt, D-38116 Braunschweig, Germany]{Physikalisch-Technische Bundesanstalt (PTB)}

\title
{Interference of Binding Protein\\
 with Serum Growth Hormone Measurements}

\begin{document}

\abstract{\noindent Measurement of serum growth hormone by mass spectrometry is demonstrated to be unaffected by interferences with growth hormone binding protein as frequently encountered with antibody-based routine test methods and provides an alternative approach, therefore, to acquisition of accurate results.}

%  Introduction              

\section{Introduction}
Interference by growth hormone binding protein (GHBP) is recognized as one of the principal sources of biases and discrepancies between laboratory results for serum growth hormone (GH), leading to misdiagnosis and inappropriate treatment of patients\cite{Wieringa2014}. Incomplete recovery of GH by antibody-based assays  is attributed to a significant extent to GH-GHBP complex formation, competing with antibody binding, and preventing, in this way, the molecule from being detected.\par
In recent years, mass spectrometry in general has emerged as an alternative approach to the quantification of targeted proteins from complex biological matrices. The potential of mass-spectrometry based GH measurement\cite{Arsene2010,Arsene2012,Wagner2014} has been demonstrated in a study about cut-off values in diagnosis of GH deficiency\cite{Wagner2014}. Mass spectrometry, as selecting the targeted molecule (or collision fragment) by its mass, does not depend on the recognition of any epitope on GH, as antibodies do. Any complexes that may be present in the sample can reasonably be assumed to be dissociated during the sample clean-up and denaturing steps involved. Proteolytic fragments used for quantification are selected so as to uniquely code for GH while excluding fragments resulting from matrix proteins. Considering this, it is obvious that interference by matrix proteins such as GHBP is expected to be negligible, if present at all, with mass-spectrometric GH assays. \par The data presented here are intended to highlight this fact, rendering mass spectrometry a complementary method or alternative, respectively, to antibody-based (ligand-binding) method principles.

% Experimantal        

\section{Experimental} 
\subsection{Samples and Reagents}
Samples were prepared by adding of GH and growth hormone receptor (GHR Fc chimera) to aliquots of GH-depleted serum. The GH-level was uniformly 9~ng/mL with all of them, whereas the receptor was added to final concentrations of 50, 100, 200 and 500~ng/mL.\par
In detail: The appropriate aliquot of a GH-stock solution (1.7~ng/$\mu$L, 50\% acetonitrile, 50~mM acetic acid) was added to 135~$\mu$L of a buffer solution consisting of Na-EDTA (5~mM), KH\ce{_2}PO\ce{_4} (20~mM) and 0.5\% bovine serum albumin in aqua. Following lyophilization, receptor solution was added, prior to collecting the material with depleted serum (6~mL). The amount of receptor solution was that needed to obtain the target level of GHR Fc chimera with that sample. Samples were then incubated for 1~h at room temperature and stored at -80~$^\circ$C until analysis.\par
Recombinant 22~kDa-GH (WHO~2.~IS~98/574) was used as the GH-material. GHR Fc chimera was sourced from R\&D Systems, Minneapolis, USA. The depleted serum had been obtained from SCIPAC, Sittingbourne, UK, with a specification of $<$0.1~ng/mL GH.

\subsection{Methods}

{\bf \em Mass Spectrometry}: Measurements were performed on a 4000 Q~Trap LC/MS-MS system (Applied Biosystems) as detailed elsewhere\cite{Wagner2014}. Briefly, a 200~$\mu$L aliquot of each serum sample was spiked with isotope labeled GH as internal standard and, after equilibration, digested with trypsin. Then, GH-cleavage fragments T6 and T12 were isolated by collecting the corresponding fractions from reversed phase chromatography and subsequent cation-exchange chromatography. These fractions were analyzed by (RP-) LC/MS-MS. The GH concentration was calculated based on acquired mass-spectrometric signals specific to T6 and T12, respectively. Coefficients of variation (CVs) have been reported to be 3.5\% (if using T6)\cite {Arsene2012} and 2.4\% (using T12).\cite{Arsene2010}\par

\noindent {\bf \em Antibody-based assays:} Immulite 2000 (Siemens \rule{3pt}{0pt} Healthcare \rule{3pt}{0pt} Diagnostics,  Erlangen,   Germany)\cite{Immulite2014} and iSYS (Immunodiagnostic Systems GmbH, Frankfurt am Main, Germany)\cite{iSYS2014} were used according to the current versions of the manufacturers' protocols. The limits of quantification were specified by the manufacturers as 0.01~ng/mL and 0.05~ng/mL (Immulite/iSYS). The specified intra- and interassay CVs were 3.5\% and 6.5\% at 2.6~ng/mL, respectively, (Immulite), whereas the CVs were specified as 4.82\% at 2.26 ng/mL with the iSYS assay\cite{Manolopoulou2012}. Both assays were calibrated against 22 kDa GH (WHO~2.~IS~98/574). 50~$\mu$L aliquots  of each of the serum samples were analyzed with both of the assays, Immulite and iSYS.

% Results and Discussion

\section{Result and Discussion}

\begin{figure}[H]
\includegraphics[width=1\textwidth]{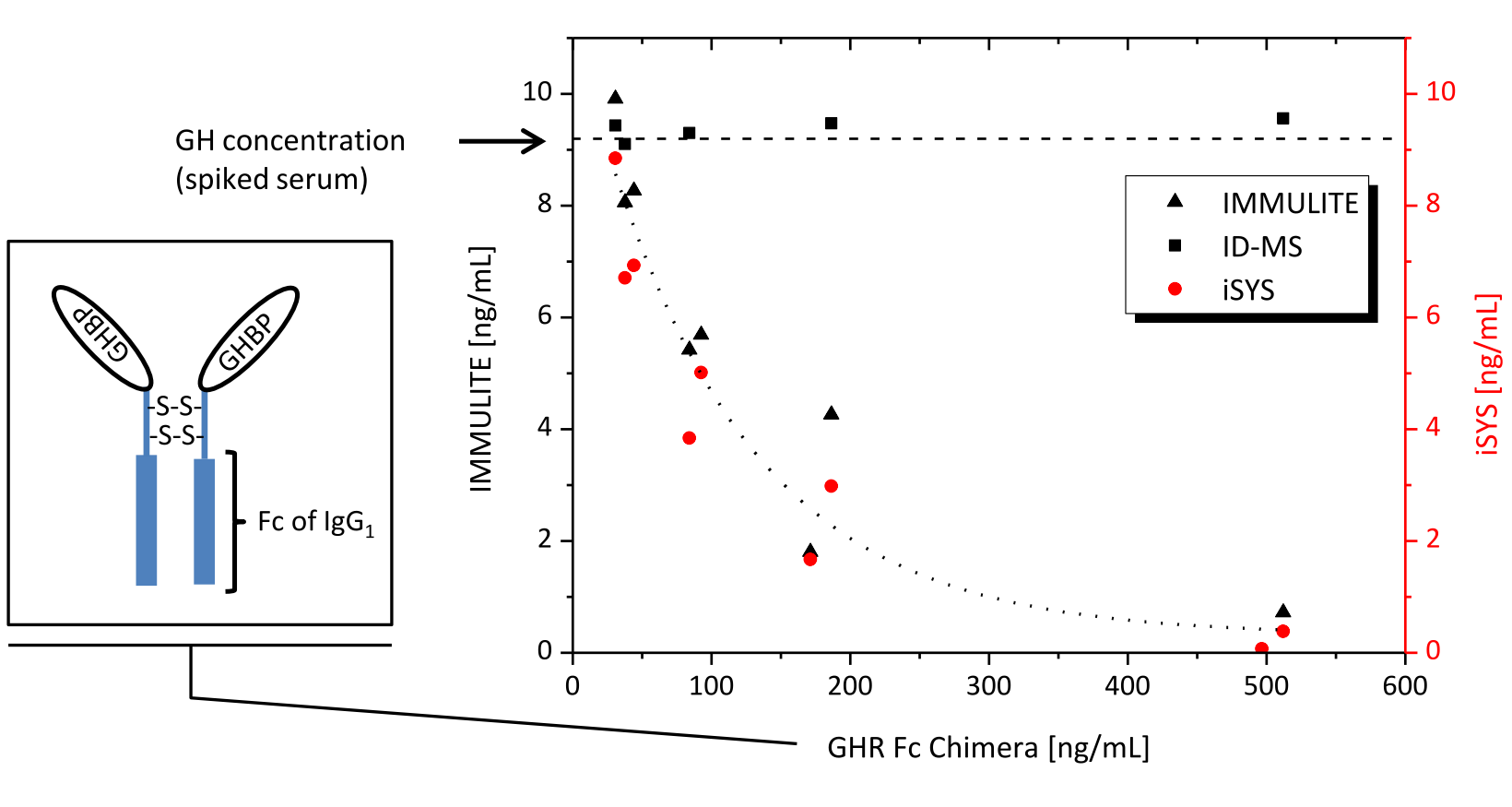}
\caption{Dependence of serum growth hormone (GH) assay results on binding protein (GHBP) level. Two commonly used antibody-based assays (IMMULITE, black triangles and iSYS, red circles) are compared to mass spectrometry (ID-MS, black squares). A GHR Fc chimera was spiked to serum so as to mimic increasing concentrations of the binding protein. The dashed horizontal line indicates the known concentration of 22 kDa-GH. (ID-MS data shown are avarages from T6-based and T12-based GH measurement.)}\label{GHBPinterference}
\end{figure}

Mass-spectrometric GH quantification, as expected, is accurate and unaffected by the amount of growth hormone receptor (GHR) added to the serum, while the outcome of both the antibody-based assays is significantly influenced by presence of GHR, as reflected by the results shown in Figure \ref{GHBPinterference}. The curve progression is in line with the assumption that the antibody and receptor are competing for the same topological feature (epitope) on the surface of the GH molecule, such that increasing concentrations of GHR turn equilibria in favor receptor binding, and reducing, accordingly, the recovery of the hormone. Assuming, the model GHR used in this study, closely reflects the behavior of the binding protein in real samples, it may be concluded that reliable measurement of GH with these assays requires correction for the amounts of binding protein not only in the sample, but also in the (serum) calibrator used with that assay.\par

It should be noted that, mass spectrometry is accurate as discussed here to the extent that the total amount of the hormone, is the measurand referred to, regardless if present free, bound to other proteins, dimerized or even in higher aggregation state. Although this apparently is present consensus, alternative definition of measurand has been considered e.g. by Frystyk et al.\cite{Frystyk2008} who separate 'free' GH from the GH in complex with the binding protein by ultrafiltration prior to antibody-based measurement. Using this method, the authors found inverse correlation of free GH concentration with binding protein levels, which is matching well with the influence of binding protein on recovery by antibody-based assays, as shown here.

%   Figure: GHBP-interference  

\rule{0pt}{60pt}
\bibliography{GHBPinterference}
\end{document}